\documentclass[USenglish]{article}	

\usepackage[utf8]{inputenc}				
\usepackage[big,online]{dgruyter}	
\usepackage{lmodern} 
\usepackage{microtype}
\usepackage[numbers,square,sort&compress]{natbib}
\usepackage{xcolor}

\theoremstyle{dgthm}

\theoremstyle{dgdef}

\usepackage[normalem]{ulem}
\begin{document}

	\articletype{Research Article}
	\received{Month	DD, YYYY}
	\revised{Month	DD, YYYY}
  \accepted{Month	DD, YYYY}
  \journalname{De~Gruyter~Journal}
  \journalyear{YYYY}
  \journalvolume{XX}
  \journalissue{X}
  \startpage{1}
  \aop
  \DOI{10.1515/sample-YYYY-XXXX}

\title{Dimensional \textcolor{black}{confinement} and waveguide effect of Dyakonov surface waves in twisted confined media}
\runningtitle{Dyakonov surface waves in twisted confined media}

\author[1]{D.~A.~Chermoshentsev}
\author[2]{E.~V.~Anikin}
\author*[2]{S.~A.~Dyakov} 
\author[2]{N.~A.~Gippius}
\runningauthor{D.~A.~Chermoshentsev, et al.}
\affil[1]{\protect\raggedright 
Skolkovo Institute of Science and Technology, Moscow Region, Russia and Moscow Institute of Physics and Technology, Moscow Region, Russia and Russian Quantum Center, Moscow Region, Russia}
\affil[2]{\protect\raggedright 
Skolkovo Institute of Science and Technology, Moscow Region, Russia, e-mail: s.dyakov@skoltech.ru}
	
	
\abstract{We theoretically study Dyakonov surface waveguide modes that propagate along the planar strip interfacial waveguide between two uniaxial dielectrics. We demonstrate that due to the one-dimensional electromagnetic confinement, Dyakonov surface waveguide modes can propagate in the directions that are forbidden for the classical Dyakonov surface waves at the infinite interface. We show that this situation is similar to a waveguide effect and formulate the resonance conditions at which Dyakonov surface waveguide modes exist. \textcolor{black}{We demonstrate that the propagation of such modes without losses is possible.} We also consider a case of two-dimensional confinement, where the interface between two anisotropic dielectrics is bounded in both orthogonal directions. We show that such a structure  supports Dyakonov surface cavity modes. Analytical results are confirmed by comparing with full-wave solutions of Maxwell's equations. We believe that our work paves the way towards new insights in the field of surface waves in anisotropic media.}


\maketitle

\section{Introduction} 

Surface electromagnetic waves, propagating along the interface of two dissimilar media, is the subject of extensive research during the last decades because they represent one of the fundamental concepts of nanophotonics. Understanding the optical properties of surface waves is of great importance for realizing their practical application. 

There are several types of surface waves that differ in a material type, a domain of existence, propagation constant, decay profile, etc. Among different types of surface waves, there are surface plasmon-polariton at a metal-dielectric interface \cite{Raether1988}, Tamm surface states at a photonic crystal boundary \cite{Malkova2006a,vinogradov10,dyakov2012surface}, surface solitons at a nonlinear interface \cite{kartashov2006surface} and many others.  

Another family of surface waves is Dyakonov surface waves (DSW) which exist at the interface of two media at least one of which is anisotropic as predicted in 1988 in Ref.\,[\!\!\citenum{Dyakonov1988}]. In this pioneering work the first medium was considered as isotropic dielectric with the refractive index $n_{m}$, while the second medium was anisotropic uniaxial dielectric with the refractive indices $n_{o}$ and $n_{e}$ and an optical axis is parallel to the interface. It has been shown that Dyakonov surface waves exist in such a system if the condition
\begin{equation}
n_{o}<n_{m}<n_{e}.
\label{eq:Dyakonov condition}
\end{equation}
is satisfied.
In 1998 Walker et al. extended the theory of Dyakonov surface waves to the case of biaxial medium \cite{Walker1998} with refractive indices $n_{x}<n_{y}<n_{z}$. In isotropic/biaxial system the condition (\ref{eq:Dyakonov condition}) transforms into
\begin{equation}
n_{x}<n_{y}<n_{m}<n_{z}
\label{eq:Dyakonov biaxial condition}
\end{equation}
Later, different combinations of isotropic, uniaxial, biaxial and chiral materials have been demonstrated to support DSWs \cite{Narimanov2018,Lakhtakia2007,Gao:09,repan2020wave, zhang2020unusual, karpov2019dyakonov,fedorin2019dyakonov}.

A narrow range of propagation angles makes the experimental observation of Dyakonov surface waves rather complicated \cite{Takayama2008}. As a result, the first detection of these waves has been demonstrated only in 2009 \cite{Takayama2009}. The authors used Otto-Kretchmann configuration to observe Dyakonov surface states at the interface of biaxial crystal and isotropic liquid. Another perspective approach to obtain Dyakonov-like surface waves experimentally is the usage of partnering thin films between anisotropic and isotropic media \cite{Takayama2014}. In such systems, the direction of hybrid Dyakonov-guided modes propagation can be controlled by changing the isotropic medium's refractive index. The results presented in Ref.\, \cite{Takayama2014} show that these types of waves can be used as a sensing unit. 
It has been demonstrated in a number of publications that Dyakonov surface waves can exist at the interface of isotropic materials and materials with artificially designed shape anisotropy \cite{Chiadini2016, Lakhtakia2007, Lakhtakia2014, Pulsifer2013, artigas2005dyakonov, jacob2008optical}. Moreover, as theoretically shown in Ref.~\,\!\!\cite{Takayama2012, Takayama2012coupling}, in the metamaterial composed of alternating layers of metals and dielectric, exotic types of surface waves such as Dyakonov plasmons and hybrid plasmons can appear. In such structures, the angular range of existence of Dyakonov surface waves can be extended up to $\Delta\phi \sim 65^{\circ}$. 

Recently, in 2019, a new type of surface waves, referred to as Dyakonov-Voigt surface waves, have been theoretically demonstrated at the interface of isotropic and uniaxial materials \cite{Mackay2019}. Unlike conventional Dyakonov surface waves, Dyakonov-Voigt surface waves decay as the product of a linear and an exponential function of the distance from the interface in the anisotropic medium \cite{Lakhtakia2020, Lakhtakia2020a,zhou2020theory}. In contrast to Dyakonov surface waves, Dyakonov-Voigt surface waves propagate only in one direction in each quadrant of the interface plane.

Like \textcolor{black}{in case of many} other surface waves, the feasibility of practical use of Dyakonov surface waves depends ultimately on whether they can exist in resonator structures of finite size \textcolor{black}{and whether they can propagate without radiative losses}. In Ref.~\,\!\!\cite{Kajorndejnukul2019} it has been shown that the Dyakonov surface waves can be conformally transformed into the bound states of cylindrical metamaterials. Dyakonov-like surface waves have been also theoretically predicted in anisotropic cylindrical waveguides \cite{Golenitskii2019}. \textcolor{black}{Due to bending of the waveguide boundary, such modes have inevitable radiative losses.}

This paper is devoted to the theoretical study of Dyakonov-like surface states at a \textit{flat} interface confined in one or two dimensions. We consider two anisotropic uniaxial lossless dielectrics \textcolor{black}{twisted in such a way that their  optical axes form an angle of $\alpha=90^{\circ}$ to each other} and are parallel to the interface plane. \textcolor{black}{We study Dyakonov surface waveguide modes in the case of one-dimensional electromagnetic confinement and show that the propagation of such modes without radiative losses is possible. We also introduce the concept of Dyakonov surface cavity modes in the case of two-dimensional confinement.} 

\section{Interface of two uniaxial crystals}\label{Sec:Infinite}
\begin{figure*}[t!]
    \centering
    \includegraphics[width=0.9\textwidth]{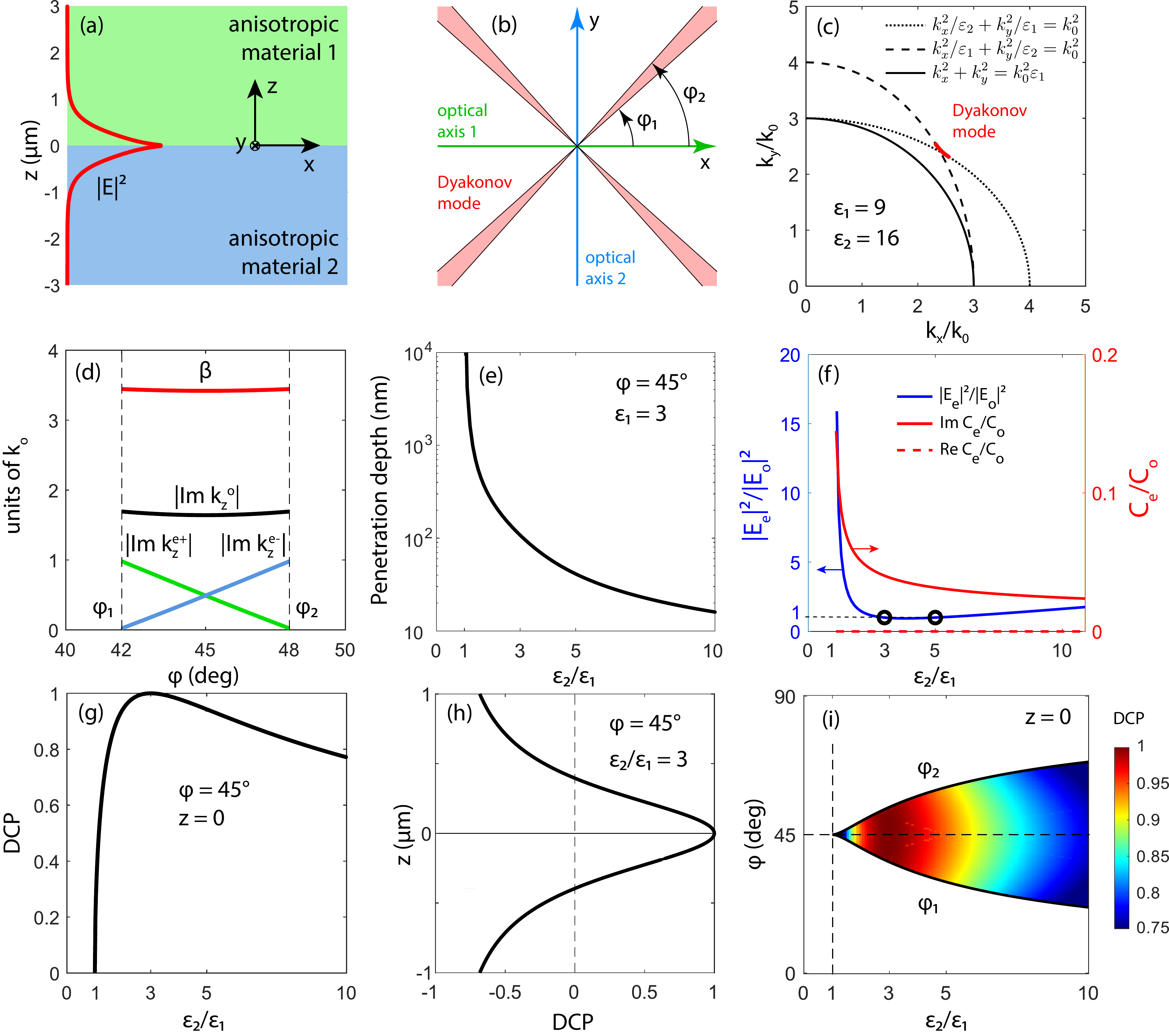}
    \caption{(Color online) (a) The interface between two anisotropic materials. Red lines show the profile of electric field intensity. (b) Top view of the interface. Red shaded regions show the propagation cones of DSWs. Angles $\varphi_1$ and $\varphi_2$ are the limits of the $\varphi$-range of existence of DSWs. (c) Ordinary wave (solid black line) and extraordinary waves in anisotropic materials 1 and 2 (dashed and dotted lines) as well as DSW (red line) in reciprocal space. (d) Azimuthal angle dependencies of propagation constant $\beta$ of DSW and the absolute value of imaginary parts of $z$-projections of wavevectors of ordinary and extraordinary waves in anisotropic materials 1 and 2 with $\varepsilon_1 = 9$ and $\varepsilon_2 = 16$. Black dashed lines bound the $\varphi$-range of the existence of DSWs.  \textcolor{black}{(e) The penetration depth of DSW into upper and lower anisotropic media at $\varphi=45^\circ$.} (f) The ratio of electric field intensities of ordinary and extraordinary waves which form DSW (blue line) and the ratio of coefficients $C_e/C_o$ (red lines) as functions of anisotropy factor $\gamma=\varepsilon_2/\varepsilon_1$. (g) Degree of circular polarization (DCP) of DSW at $z=0$ propagating at $\varphi=45^\circ$ as a function of anisotropy factor $\gamma$.(h) DCP of DSW propagating at $\varphi=45^\circ$ as a function of coordinate $z$ at $\varepsilon_2/\varepsilon_1=3$. (i) Thick black lines denote the limits of the $\varphi$-range existence of DSWs as functions of anisotropy factor, $\varphi_2(\varepsilon_2/\varepsilon_1)$ and $\varphi_1(\varepsilon_2/\varepsilon_1)$. \textcolor{black}{Colormap} shows the anisotropy factor and azimuthal angle dependence of the DCP of DSW. Colorscale is shown on the right.}
\label{fig:infinite_space}
\end{figure*}

We start our discussion by considering a flat infinite interface between two \textcolor{black}{twisted} semi-infinite anisotropic uniaxial media shown in Fig.~\ref{fig:infinite_space}a. \textcolor{black}{We assume that the optical axes of the upper and lower media are directed along the $x$ and $y$ coordinate axes. As shown in Refs.\,\cite{averkiev1990electromagnetic, Darinskii2001, Furs2005}, such a configuration supports DSWs when anisotropic media are optically positive, i.e. the condition\,\eqref{eq:Dyakonov condition} is satisfied. In Refs.\,\cite{Furs2005a, Nelatury2007, PoloJr.2007, Alshits2002} the problem of DSWs have been generalized to the case of two biaxial crystals. In this section we once again describe some of the key points of DSWs in the uniaxial/uniaxial configuration which are crucial for understanding the properties of Dyakonov-like surface waves in confined media.} We denote the dielectric permittivity tensor of the upper half-space as $\hat{\epsilon}_{\mathrm{up}} =$ diag($\epsilon_{2}, \epsilon_{1}, \epsilon_{1}$) and of the lower half-space as $\hat{\epsilon}_{\mathrm{low}} =$ diag($\epsilon_{1}, \epsilon_2, \epsilon_{1}$). Dyakonov surface waves in such system can be obtained as a linear combination of exponentially decaying ordinary and extraordinary waves which are solutions of Maxwell's equation in each half-space:
\begin{equation}
    \begin{gathered}
        \left.\begin{matrix}
        \vec{E}_\mathrm{\mathrm{\scriptscriptstyle DSW}}^+ = C_o^+ \vec{E}_o^+ + C_e^+ \vec{E}_e^+\\
        \vec{B}_\mathrm{\mathrm{\scriptscriptstyle DSW}}^+ = C_o^+ \vec{B}_o^+ + C_e^+ \vec{B}_e^+
        \end{matrix}\right\}\quad\text{at}\quad z > 0,\\
        \left.\begin{matrix}
        \vec{E}_\mathrm{\mathrm{\scriptscriptstyle DSW}}^- = C_o^- \vec{E}_o^- + C_e^- \vec{E}_e^-\\
        \vec{B}_\mathrm{\mathrm{\scriptscriptstyle DSW}}^- = C_o^- \vec{B}_o^- + C_e^- \vec{B}_e^-
        \end{matrix}\right\}\quad\text{at}\quad z < 0.
    \end{gathered}
\end{equation}
where $\vec{E}$ and $\vec{B}$ are the electric and magnetic vectors, $C_{o}^{\pm}$ and $C_{e}^{\pm}$ are the coefficients of the linear combination and signs "+" and "-" denote upper and lower half-spaces, respectively. Using the expressions for ordinary and extraordinary waves in upper and lower anisotropic media (see Supplemental Materials) and taking into account the continuity of the in-plane field components at the interface, after algebraic manipulations one can obtain the dispersion relation of Dyakonov surface waves in the following form:
\begin{equation}
    \det\begin{pmatrix}
            0 & k_0^2\epsilon_1 - k_x^2 & k_0k_z^{o-} & k_xk_y\\
            k_0^2\epsilon_1 - k_x^2 & 0 & k_xk_y & -\epsilon_1 k_0k_z^{e-}\\
            -k_0k_z^{o+} & k_xk_y & 0 & k_0^{2}\epsilon_1 - k_y^2\\
            k_xk_y & \epsilon_1 k_0k_z^{e+} & k_0^2\epsilon_1 - k_y^2 & 0\\
        \end{pmatrix} = 0
\label{disp_eq}
\end{equation}
where $k_0=2\pi/\lambda$ is the vacuum wavenumber, $k_{x}$, $k_{y}$ and $k_{z}$ are the wavevector components, $\lambda$ is the wavelength and the $z$-components of wavevectors are:
\begin{equation}\label{kz}
    \begin{gathered}
     \left.\begin{matrix}
        k_z^{o+} = \sqrt{k_0^2\epsilon_1 - k_x^2 - k_y^2},\\
        k_z^{e+} = \sqrt{k_0^2\epsilon_2 - \gamma k_x^2 - k_y^2},\\
        \end{matrix}\right\}\quad\text{at}\quad z > 0,\\
     \left.\begin{matrix}
       k_z^{o-} = -\sqrt{k_0^2\epsilon_1 - k_x^2 - k_y^2}\\
       k_z^{e-} = -\sqrt{k_0^2\epsilon_2 - k_x^2 - \gamma k_y^2}
        \end{matrix}\right\}\quad\text{at}\quad z < 0,\\
    \end{gathered}
\end{equation}
where $\gamma=\epsilon_{2}/\epsilon_{1}$ is the anisotropy factor. In Eq.\,\eqref{kz}, all $k_z$ are purely imaginary and the signs of square roots are chosen in such a way that the solution decays with distance from the interface in upper and lower half-spaces.

The numerical solution of the equation (\ref{disp_eq}) for the Dyakonov wave is represented in Fig.\ref{fig:infinite_space}c by the red curve. One can see that the DSW is located near the intersection of the dispersion curves of extraordinary waves with $k_z=0$ in upper and lower half-spaces. The ($k_x, k_y$)-range of existence of DSW determines the narrow domain of azimuthal angle $\varphi$ near the bisector between the crystals' optical axes where the DSW can propagate (Fig.\,\ref{fig:infinite_space}b). This is in agreement with the results reported in \textcolor{black}{ Refs.\,[\citenum{averkiev1990electromagnetic,Furs2005,Takayama2011}]} for the two symmetrical uniaxial anisotropic crystals.

The \textcolor{black}{$\varphi-$}angular dependencies of $|\mathrm{Im}(k^{o,e}_{z})|$ and the propagation constant \textcolor{black}{ $\beta(\varphi)=\sqrt{k_x^2+k_y^2}$} for permittivities $\epsilon_{1}=9$ and $\epsilon_{2}=16$ are presented in Fig.\ref{fig:infinite_space}d. One can see that the extraordinary wave decays slower than the ordinary wave. In the cut-off points $\varphi_{1}$ and $\varphi_{2}$, the imaginary part of $|\mathrm{Im}(k^{e}_{z})|$ turns to zero and the solution is no longer localized near the interface. It is worth noting that at $\varphi=45^{\circ}$, the absolute values of $k^{e+}_z$ and $k^{e-}_z$ are the same, and the DSW decays to the upper and lower half-spaces equally.

To estimate the partial contributions of ordinary and extraordinary waves to the \textcolor{black}{DSW} we calculate the ratio of the coefficients $C_e/C_o$ as well as the ratio of the electric field intensities $|E_e|^2/|E_o|^2$ in the most symmetric case $\varphi=45^\circ$ when $C_{o,e}^+=C_{o,e}^-$ and $|E_{o,e}^+|^2=|E_{o,e}^-|^2$. Fig.\,\ref{fig:infinite_space}\textcolor{black}{f} demonstrates that at $\gamma\lesssim2$ the contribution of extraordinary wave is dominant. \textcolor{black}{Knowledge of the partial contributions of the ordinary and extraordinary waves to the DSW and their decay constants $\mathrm{Im} k_z$ (Fig.\,\ref{fig:infinite_space}d) enables us to find the resulting penetration depth of DSW into the upper and lower anisotropic dielectrics (Fig.\,\ref{fig:infinite_space}e). One can see in Fig.\,\ref{fig:infinite_space}e that with increase of the anisotropy factor, the DSW becomes more localized at the interface.}

Dispersion relation of DSW can be found analytically from Eq.\,\eqref{disp_eq} for the symmetric case of $\varphi=45^{\circ}$ ($k_{x}=k_{y}$), when the DSW propagates along the bisector. The expression for the propagation constant of DSW \textcolor{black}{in this case} reads:
\begin{equation}
    \beta(\pi/4)=k_{0} \sqrt{\frac{{\epsilon_{1}+\sqrt{(2\epsilon_{2}-\epsilon_{1})\epsilon_{1}}}}{2}}
\label{beta}
\end{equation}
Moreover, Eq.\,\eqref{disp_eq} can also be analytically solved in the cut-off points $\varphi_{1}$ and $\varphi_{2}$ of the angular domain of existence. At $\varphi=\varphi_{1}$ we get 
\begin{equation}
\label{eq:angle_kx}
        k_{x}(\varphi_1)=k_{0}\sqrt{\epsilon_{1}}\sqrt{\frac{\gamma(\gamma+2)\textcolor{black}{-}\sqrt{\gamma(\gamma^2+\gamma-1)}}{2(\gamma+1)}},
\end{equation}
\begin{equation}
\label{eq:angle_ky}
        k_{y}(\varphi_1)=k_{0}\sqrt{\epsilon_{1}}\sqrt{\frac{\gamma^2\textcolor{black}{+}\sqrt{\gamma(\gamma^2+\gamma-1)}}{2\gamma(\gamma+1)}}.
\end{equation}
This gives us the analytical expression for the cut-off angle $\varphi_1$:
\begin{equation}
\label{eq:bound_tg1}
       \tan{\varphi_{1}}=\frac{k_{y}(\varphi_1)}{k_{x}(\varphi_1)}=\sqrt{\frac{\gamma^2\textcolor{black}{+}\sqrt{\gamma(\gamma^2+\gamma-1)}}{\gamma^2(\gamma+2)\textcolor{black}{-}\gamma\sqrt{\gamma(\gamma^2+\gamma-1)}}}.
\end{equation}
Owing to the symmetry, the second cut-off angle $\varphi_{2}$ can be determined by swapping $k_{x}$ and $k_{y}$. This \textcolor{black}{brings} us to a simple relation $\tan\varphi_{2}=\cot\varphi_{1}$ (Fig.\ref{fig:infinite_space}b).
One can see from Eq.\,\ref{eq:bound_tg1} that the cut-off angles depend only on the anisotropy factor $\gamma=\epsilon_{2}/\epsilon_{1}$ and does not depend on the values of $\epsilon_{1}$ and $\epsilon_{2}$. One can also notice that the higher the anisotropy factor, the larger the angular domain of existence (Fig.\ref{fig:infinite_space}i). Due to the structure symmetry, the DSW can propagate in identical angular domains rotated relative to the OZ axis by $\pi/2$. (Fig.\ref{fig:infinite_space}b).

Like many surface waves, DSW are \textcolor{black}{elliptically} polarized. However, the degree of circular polarization (DCP) depends on the anisotropy factor $\gamma$, on the azimuthal angle of propagation $\varphi$ and on the coordinate $z$ where the electric field is considered. For the most symmetric case  (\textcolor{black}{$\varphi=45^\circ$}, $z=0$) the $\gamma$-dependence of the DCP can be expressed analytically:
\begin{equation}
    \label{eq:dcp}
    \mathrm{DCP} = 2\sqrt{2}\frac{\sqrt{\gamma-1}}{\gamma+1}.
\end{equation}
From Eq.\,\eqref{eq:dcp} we can see that at $\gamma=3$ the DCP equals to 1 (Fig.\ref{fig:infinite_space}g) which corresponds to a purely circularly polarized field. In the limit of a low anisotropy, the DCP goes to 0 and the DSW becomes almost linearly polarized. By means of full-wave electromagnetic simulations made by scattering matrix method \cite{li2003fourier, weiss2009matched, Tikhodeev2002b} we calculate the DCP in less symmetric case when $z\ne0$. We obtain that with distance from the interface, the orientation of polarization cones in DSW changes, \textcolor{black}{wherein} the DCP decreases and changes its sign (Fig.\,\ref{fig:infinite_space}h). We also vary the azimuthal angle $\varphi$ (Fig.\,\ref{fig:infinite_space}i) and find that the $\varphi$-dependence of the DCP is weak, however for all $\varphi\ne45^\circ$ DCP$<1$. 
\section{Reflection from boundary}\label{sec:reflection}

\begin{figure*}[t!]
    \centering
    \includegraphics[width=1\textwidth]{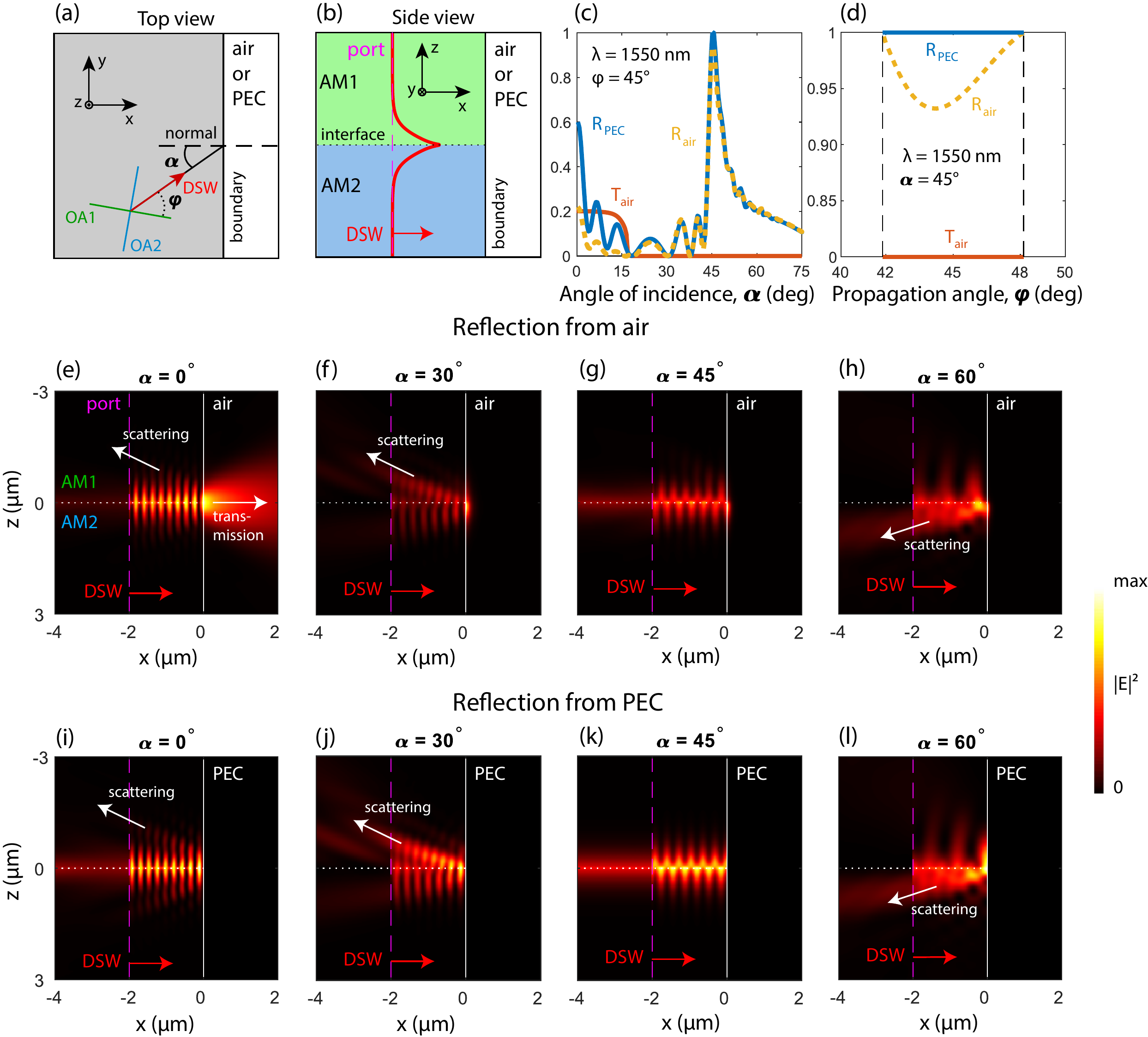}
    \caption{(Color online) Top view (a) and side view (b) of the interface between two anisotropic materials bounded by air or a perfect electric conductor (PEC) half-space on the right. In panel (b) the red line schematically shows the Dyakonov surface wave, which is induced by the port denoted by dashed magenta line, and then hits the boundary. (c) \textcolor{black}{and (d) The $\alpha$- and $\varphi$-}angular dependence\textcolor{black}{s} of specular reflection and transmission in such configuration.  \textcolor{black}{In (d) the black dashed lines bound the $\varphi$-range of the existence of DSWs in the infinite interface.} Vertical cross-section of electric field intensity when Dyakonov surface wave hits the air boundary \textcolor{black}{(e--h)} or the PEC boundary \textcolor{black}{(i--l)} calculated at different incident angles $\alpha$. Arrows show the direction of scattering and transmission. All simulations are made for $\lambda=1550$ nm, $\epsilon_{1}=9$ and $\epsilon_{2}=16$.  Colorscale is shown on the right.}
    \label{fig:reflection_case}
\end{figure*}

Before \textcolor{black}{moving on to exploring the dimensional confinement of Dyakonov-like surface states it is important} to analyze the scattering of a DSW on a single boundary perpendicular to the interface plane along which the DSW propagates. The results obtained in the previous section provide a possibility to perform this analysis.

We consider a DSW propagating along the interface \textcolor{black}{at a varying azimuthal propagation angle $\varphi$} and hitting the boundary at a varying angle of incidence $\alpha$ (Fig. \ref{fig:reflection_case}a). \textcolor{black}{In this section we use a new coordinate system where the boundary is parallel to $y$-axis and the angle between the optical axes and coordinate axes equals to $\varphi-\alpha$.} We consider two cases when the boundary separates anisotropic materials from (i) air and (ii) perfect electric conductor (PEC). In reflection, the $y$-component of the wavevector $k_{y}(\alpha)$ is conserved and, hence, can be described by the propagation constant $\beta$ in the following way:
\textcolor{black}{
\begin{equation}
    k_{y}(\alpha,\varphi)=\beta(\varphi)\sin{(\alpha+\varphi-\pi/4)}.
\label{reflection_perp}
\end{equation}}
It enables us to reduce the 3D scattering problem to a 2D problem in the XZ plane with a fixed out-of-plane wavevector component $k_{y}(\alpha,\varphi)$ and the corresponding orientation of optical axes of anisotropic materials (Fig. \ref{fig:reflection_case}b). We perform the corresponding electromagnetic simulations of DSW scattering on the boundary \textcolor{black}{using the Finite Element Method (FEM)} in COMSOL Multiphysics.



\textcolor{black}{Firstly, we explore the most symmetric configuration fixing the azimuthal propagation angle at $\varphi=45^\circ$ and  calculate the specular reflection and transmission coefficients at varying $\alpha$ for the air or PEC boundaries (Fig. \ref{fig:reflection_case}c)}. Since both incident and reflected DSWs can only propagate in the limited domain near the bisector between the optical axes, reflection of this mode without significant scattering losses can occur only at the angle $\alpha$ close to 45$^\circ$. In the case of air boundary, the transmission turns to zero at $\alpha>17.5^\circ$ which is due to the total internal reflection at incident angles exceeding critical angle. Please note that we do not consider the transmission to PEC because it is zero by definition.

\textcolor{black}{Secondly, we consider the case the maximal reflection fixing the incident angle $\alpha$ at 45$^{\circ}$ and varying the azimuthal propagation angle $\varphi$ within the $\varphi$-range of DSW existence $[\varphi_1, \varphi_2]$ (Fig.\,\ref{fig:reflection_case}d). Please note that in the considered case, the crystals' orientation is fixed. One can see in Fig.\,\ref{fig:reflection_case}d that for the PEC boundary, the reflectance of DSW  is independent of the $\varphi$ and equals to 1. In the case of the air boundary, the reflection coefficient $R<1$ for all $\varphi$ except for the cut-off points $\varphi_1$ and $\varphi_2$, where the DSW transforms into the extraordinary propagating wave of one of the two half-spaces. The asymmetry of the reflection coefficient profile results from i) the non-identity of the upper and lower half-spaces in the presence of the boundary; ii) assymetry of the decay profiles of DSW in the upper and lower half-spaces at $\varphi\ne 45^\circ$ (see Fig.\,\ref{fig:infinite_space}d). }

Figs.\,\ref{fig:reflection_case}\textcolor{black}{e--l} show the profiles of the period average electric field intensity of the DSW \textcolor{black}{being} produced by the port and falling on the boundary at different incident angles $\alpha$ and \textcolor{black}{at a fixed azimuthal propagation angle $\varphi=45^\circ$}. One can see from Fig.\,\ref{fig:reflection_case}\textcolor{black}{e} that at $\alpha=0^\circ$ the incident wave is partially reflected, scattered and transmitted to the air. Intensity modulation between the port and the boundary is due to the interference of incident and reflected DSWs. At $\alpha=45^\circ$ (Fig.\,\ref{fig:reflection_case}\textcolor{black}{g}) the DSW is reflected back to the interface. In the case of air boundary there is a slight scattering which occurs only to the incident side \textcolor{black}{as there is a} total internal reflection. The specularly reflected DSW propagates perpendicularly to the incident DSW, along the second bisector between two orthogonal optical axes. In the less symmetric cases of $\alpha=30^\circ$ and $60^\circ$  the wave is scattered out to the anisotropic media almost completely. Similar behaviour of DSW is observed for the case of PEC boundary with the exception of the fact that at $\alpha=45^\circ$ no scattering occurs.

\textcolor{black}{We would like to emphasize once again that a non-perfect reflection at $\alpha=45^\circ$ from the air boundary occurs due to the coupling of DSW to the propagating modes of the upper and lower half-spaces, but not due to the transmission to the air.}


\section{One-dimensional confinement}
\label{sec:1d confinement}

\begin{figure*}[t!]
    \centering
    \includegraphics[width=1\textwidth]{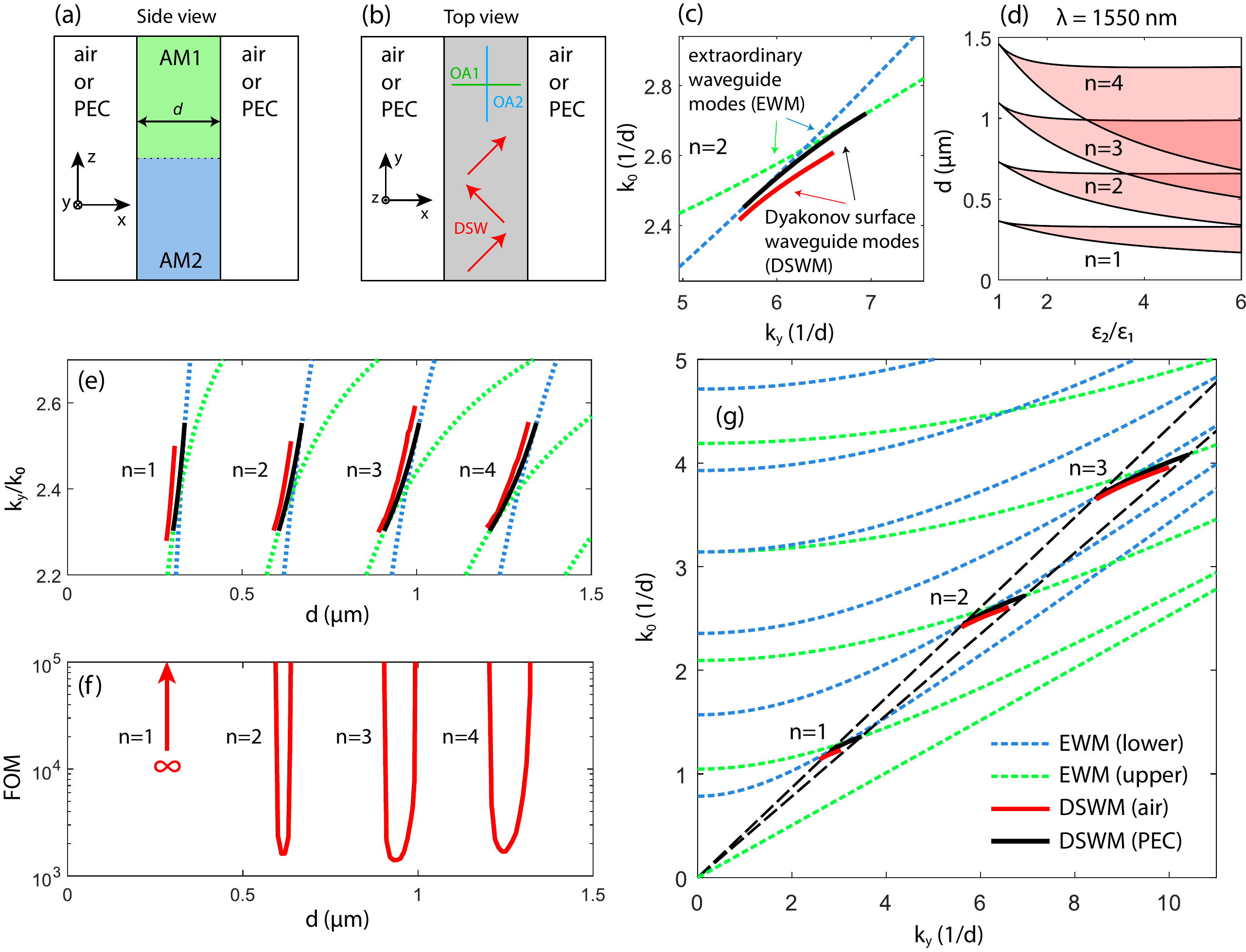}
    \caption{(Color online) Side view (a) and top view (b) of the interface between two anisotropic materials bounded by air or PEC half-spaces from left and right. Optical axes of anisotropic materials are parallel to coordinate axes as shown by the green and blue lines in panel (b). DSWM in such configuration is a  superposition of DSWs reflecting from both sides of the boundary at the angle $\alpha$ \textcolor{black}{close to} 45$^\circ$ as is shown in panel (b) by red arrows. (c), (e) and (g): Extraordinary waveguide modes (EWM) of the upper and lower anisotropic slabs (dashed green and blue lines) with PEC boundaries and DSWMs in the case of a PEC (solid black line) or air (solid red line) \textcolor{black}{boundary}. (d): Range of the waveguide width, $d$, in which  DSWMs exist. (f): Figure of merit (FOM) calculated for the case of air \textcolor{black}{boundary. All calculations are made for $\varepsilon_1=9$, $\varepsilon_2=16$ except for panel (d).}}
    \label{fig:zigzag_case}
\end{figure*}
\begin{figure*}[t!]
    \centering
    \includegraphics[width=0.9\textwidth]{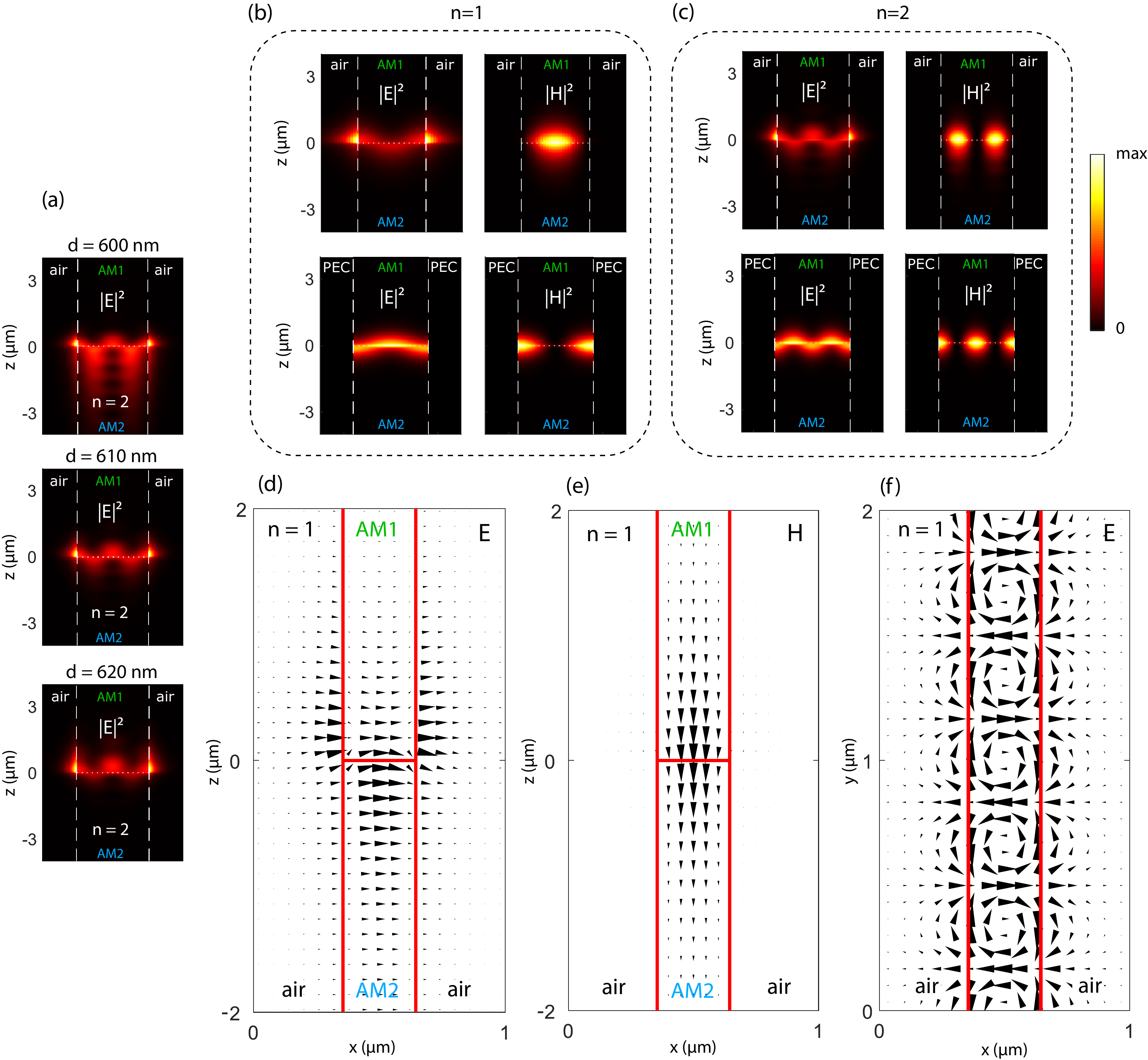}
    \caption{(Color online) (a): Electric field intensity profiles in $xz$ cross-section in the 2-nd order DSWM calculated for different waveguide widths.
    (b),(c): Electric field intensity profiles in $xz$ cross section in the 1-st order and the 2-nd order DSWMs calculated for the air and PEC \textcolor{black}{boundaries}. Calculations are made for  $d=290$\,nm, $k_y=9.429$\,$\mu$m$^{-1}$ (air, $n=1$); $d=320$\,nm, $k_y=9.786$\,$\mu$m$^{-1}$ (PEC, $n=1$); $d=610$\,nm, $k_y=9.584$\,$\mu$m$^{-1}$ (air, $n=2$); $d=640$\,nm, $k_y=9.786$\,$\mu$m$^{-1}$ (PEC, $n=2$). Colorscale for panels (a)--(c) is shown on the right. (d)--(f): Electric and magnetic vectors in $xz$ and $xy$ cross sections. The length of the triangles is proportional to the field strength at the central point of each triangle. Triangles specify the corresponding electric field direction by their orientation. White lines in (a)--(c) and red lines in (d)--(f) denote the material boundaries. All calculations are made for $\varepsilon_1=9$, $\varepsilon_2=16$,  $\lambda=1550$\,nm. }
    \label{fig:zigzag_case2}
\end{figure*}

Let us now study the system of upper and lower slabs, tangent to each other, confined between by two parallel boundaries located at $x=0$ and $x=d$ as shown in Figs.\,\ref{fig:zigzag_case}a,b. Like in the previous section, we consider the cases of air or PEC boundaries. In Fig.\,\ref{fig:reflection_case} we have demonstrated that the maximal reflectance of DSW from the boundary is reached at the incident angle $\alpha=45^{\circ}$. If the reflected wave encounters yet another boundary, parallel to the first one, then the process of multi-reflection continues and lasts until the energy is lost due to scattering. Provided that one of the optical axes is parallel to the boundary while another one is perpendicular to the boundary (Fig.\,\ref{fig:zigzag_case}b), one can expect the existence of a Dyakonov surface waveguide mode (DSWM) in the system of two anisotropic materials confined between two boundaries. While one can also consider the case of $\alpha=0$ where the refection spectrum in Fig.\,\ref{fig:reflection_case}c has a local maximum (see Supplemental Materials for details), in this section we are focused on $\alpha=45^{\circ}$.


Since such a strip waveguide has mirror symmetries $x\to -x$ and $y\to -y$, in the case of ideal reflection by PEC it is possible to treat the propagation of DSWM analytically quite easily.
If a solution for DSWs in the infinite interface
\begin{equation}
  \vec{E}_{k_x,k_y}(x,y,z) = \vec{E}_{k_x,k_y}(z) e^{ik_x x + ik_y y}  
\end{equation}
is known then the solution for DSWMs which satisfies mirror boundary conditions at $x=0$ is the following:
\begin{equation}\label{eq:zz_dyakonov}
    \vec{E}^{\mathrm{tot}}(x,y,z) = \vec{E}_{k_x,k_y}(z) e^{ik_x x + ik_y y} -  \vec{E}_{-k_x,k_y}(z) e^{-ik_x x + ik_y y}
\end{equation}
Moreover, it also satisfies mirror boundary conditions at $x = d$ provided that $k_x d = \pi n$, where $n\in\mathbb{N}$. 

These considerations enable us to find the dispersion law of DSWMs. First, we note that the dispersion law of DSW at an infinite interface in $xy$-plane can be written in a form 
\begin{equation}
    F\left(\frac{k_x}{k_0}, \frac{k_y}{k_0}\right) = 0,
\end{equation}
where a function $F$ does not depend on $k_0$ explicitly (see Eq.~\eqref{disp_eq}).
From the condition $k_x d = \pi n$, the dispersion relation of the DSWM propagating along the $y$-axis with the propagation constant $k_y$ has the following form:
\begin{equation}
    F\left(\frac{\pi n}{k_0d}, \frac{k_y}{k_0}\right) = 0,
\label{eq:dispersion_zz}
\end{equation}
where $n$ is the mode order.

The dispersion curves of DSWMs $k_0(k_y)$ calculated by Eq.\,\eqref{eq:dispersion_zz} for the case of PEC boundary and by COMSOL for the case of air boundary are shown in Figs.\,\ref{fig:zigzag_case}c,g by black and red lines respectively. For comparison, Figs.\,\ref{fig:zigzag_case}c,g also show the dispersions of extraordinary waveguide modes (EWM) of the upper and lower slabs, which are respectively TE and TM polarized because of the specific dielectric tensor orientations and PEC boundary condition. Both $k_y$ and $k_0$ are plotted in $1/d$ units, making the displayed dispersion curves universal in terms of the waveguide width $d$. One can see that the DSWMs appear near the intersection of the EWMs. We note that the dispersion curves of DSWMs have the cut-off points which originate from the angular cut-offs \footnote{Please note that the cut-off points for DSWM for the air \textcolor{black}{boundary} are determined approximately due to limitation of the computational domain size in COMSOL.} \textcolor{black}{$\varphi_1$ and $\varphi_2$} of the DSWs at the infinite interface. \textcolor{black}{Each point on the dispersion curves between the cut-offs corresponds to a certain azimuthal propagation angle $\varphi$. Please note that since the orientation of optical axes of the upper and lower waveguides are fixed, the angle $\varphi$ ultimately determines the angle of incidence $\alpha$. Hence, each point on the DSWM dispersion curves also corresponds to a certain angle of incidence $\alpha$ close to 45$^\circ$.}

The waveguide width dependence of the propagation constant $k_{y}(d)$ can also be calculated by Eq.\,\eqref{eq:dispersion_zz} for the case of PEC and is shown in Fig.\,\ref{fig:zigzag_case}e along with the same dependence calculated in COMSOL for the case of air. For comparison, $k_{y}(d)$ dependencies of extraordinary waveguide modes (EWM) of the upper and lower slabs are also shown in Fig.\,\ref{fig:zigzag_case}e. The $d$-range where DSWM can propagate is determined by the angular existence domain of the DSW at the infinite interface. With increase of the anisotropy factor, the existence range of DSWM broadens (Fig.\,\ref{fig:zigzag_case}d) and, at large anisotropy, the existence ranges of DSWMs with different mode numbers $n$ overlap.
It is worthy to note that we compared the analytical results obtained from Eq.\,\eqref{eq:dispersion_zz} for PEC with full-wave simulations made in COMSOL Multiphysics and observed an excellent agreement (not shown in Fig.\,\ref{fig:zigzag_case}).

As it has been demonstrated in Fig.\,\ref{fig:reflection_case}, the reflection of DSWMs from the PEC half-space at $\alpha=45^\circ$ is ideal, meaning that there is no scattering and, therefore, the DSWMs \textcolor{black}{between PEC boundaries} are lossless. However, in the case air half-space, even at $\alpha=45^\circ$, the reflection coefficient $R<1$. This means that in the strip waveguide surrounded by air, DSWMs can have radiative losses, which scatter out the DSWM energy to the waveguide modes of the upper and lower slabs. Like in Ref.\,[\citenum{Golenitskii2019}], to describe the radiation losses quantitatively, we calculate the following figure of merit (FOM)
\begin{equation}
    \mathrm{FOM} = \frac{\mathrm{Re}\, k_y}{\mathrm{Im}\,k_y},
\end{equation}
which has the meaning of a DSWM decay length expressed in units of the DSWM wavelength. Fig.\,\ref{fig:zigzag_case}f shows the FOM calculated in COMSOL. One can see that \textcolor{black}{for $n=2,3,4$} the FOM tends to infinity near the cut-off points while having a local minimum between them. \textcolor{black}{This behaviour of the FOM is explained by the DSW reflection profile shown in Fig.\,\ref{fig:reflection_case}d. Indeed, at the cut-off points the DSW transforms into the EWMs which have no radiative losses when reflecting from the air boundary.} We also observe that for the 1-st order DSWM ($n=1$) the FOM is infinite. To explain this \textcolor{black}{remarkable} fact, we \textcolor{black}{compare the field symmetries of the DSWMs and the waveguide modes of the upper and lower slabs (see Supplemental materials).}

Our \textcolor{black}{field simulations} reveal that for the 1-st order DSWM \textcolor{black}{there is the symmetry mismatch and the corresponding overlap integrals vanish which indicates that the coupling of the 1st order DSW mode with the slabs' waveguide modes is not possible.} As there is \textcolor{black}{also} no radiative leakage to the air (see Fig.\,\ref{fig:reflection_case}f and its discussion), we conclude that the 1-st order DSWM has no radiative losses which results in the infinite FOM. Radiative losses of higher order DSWMs are fully attributed to the coupling with the slabs' waveguide modes. \textcolor{black}{We emphasize once again the importance of the obtained result that although DSWMs are generally coupled to propagating EWMs, a symmetry-protected lossless 1st-order Dyakonov surface waveguide mode exists.}

Let us consider the field distributions in DSWMs. Cross-sectional electric field profiles of the 2-nd order DSWM in the strip waveguide surrounded by air calculated for different waveguide widths $d$ within the range of DSWM existence are shown in Fig.\,\ref{fig:zigzag_case2}a. One can see that at $d=610$\,nm the mode is localized near the interface almost equally penetrating into the upper and lower slabs. Whereas at widths close to the cut-offs, $d=600$\,nm and 620\,nm, the mode localization appears upward or downward-biased. DSWMs inherit these peculiar properties of their localization from the classical DSWs at the infinite interface. In the case of the PEC \textcolor{black}{boundary}, the waveguide widths $d_n$ corresponding to the most symmetric DSWM mode penetration into the slabs can be found from Eq.\,\eqref{eq:dispersion_zz} by setting the azimuthal propagation angle of DSW as $\varphi=45^\circ$: 
\begin{equation}
    \label{eq:symmpen}
    d_n=\pi n/k_{y}.
\end{equation}
In the case of air \textcolor{black}{boundary}, this condition will be more complex. The biasing of the DSWMs towards upper or lower slabs explains the \textcolor{black}{presence of the} local minimum in the waveguide width dependence of the FOM shown in Fig.\,\ref{fig:zigzag_case}f.

Electric and magnetic field intensity profiles of the 1-st order and the 2-nd order DSWMs are shown in Figs.\,\ref{fig:zigzag_case2}b,c for the widths $d$ such that the symmetry condition \eqref{eq:symmpen} is satisfied. For the air \textcolor{black}{boundary}, the electric field intensity profile of the $n$-th order DSWM has $n+1$ local maxima in the upper slab and $n$ local maxima in lower slab. Magnetic field intensity has $n$ local maxima in both cases. For the PEC \textcolor{black}{boundary}, the situation is different: $n$ (or $n+1$) local maxima in the upper (or lower) waveguide for electric field and $n+1$ local maxima for the magnetic field. Projections of electric and magnetic vectors directions on $xz$ and $xy$ planes are shown in Figs.\,\ref{fig:zigzag_case2}d-f for the 1-st order DSWM. One can see the periodic pattern in Fig.\,\ref{fig:zigzag_case2}f which demonstrates the propagation of DSWM along the strip waveguide. It is necessary to note here that electric and magnetic fields in DSWM are \textcolor{black}{elliptically} polarized like those in conventional DSWs \textcolor{black}{at the infinite interface}, however the orientation of polarization ellipse and the degree of circular polarization depend also on the $x$ coordinate (See Supplemental Materials for details). The examples of field distribution for cases when $d_n\neq\pi n/k_{y}$ and for different mode orders $n$ are presented in Supplemental materials.

At the end of this Section, we conclude, that the one-dimensional electromagnetic confinement makes DSWMs traveling along the direction where classical DSWs cannot propagate. Indeed, as is shown in Fig.\,\ref{fig:infinite_space}b, DSWs exist in a small angle around the bisector between optical axes of upper and lower anisotropic materials, while the DSWMs propagate along one of these optical axes. This feature distinguishes DSWMs from DSWs.


\section{Two-dimensional confinement}

\begin{figure*}[t!]
    \centering
    \includegraphics{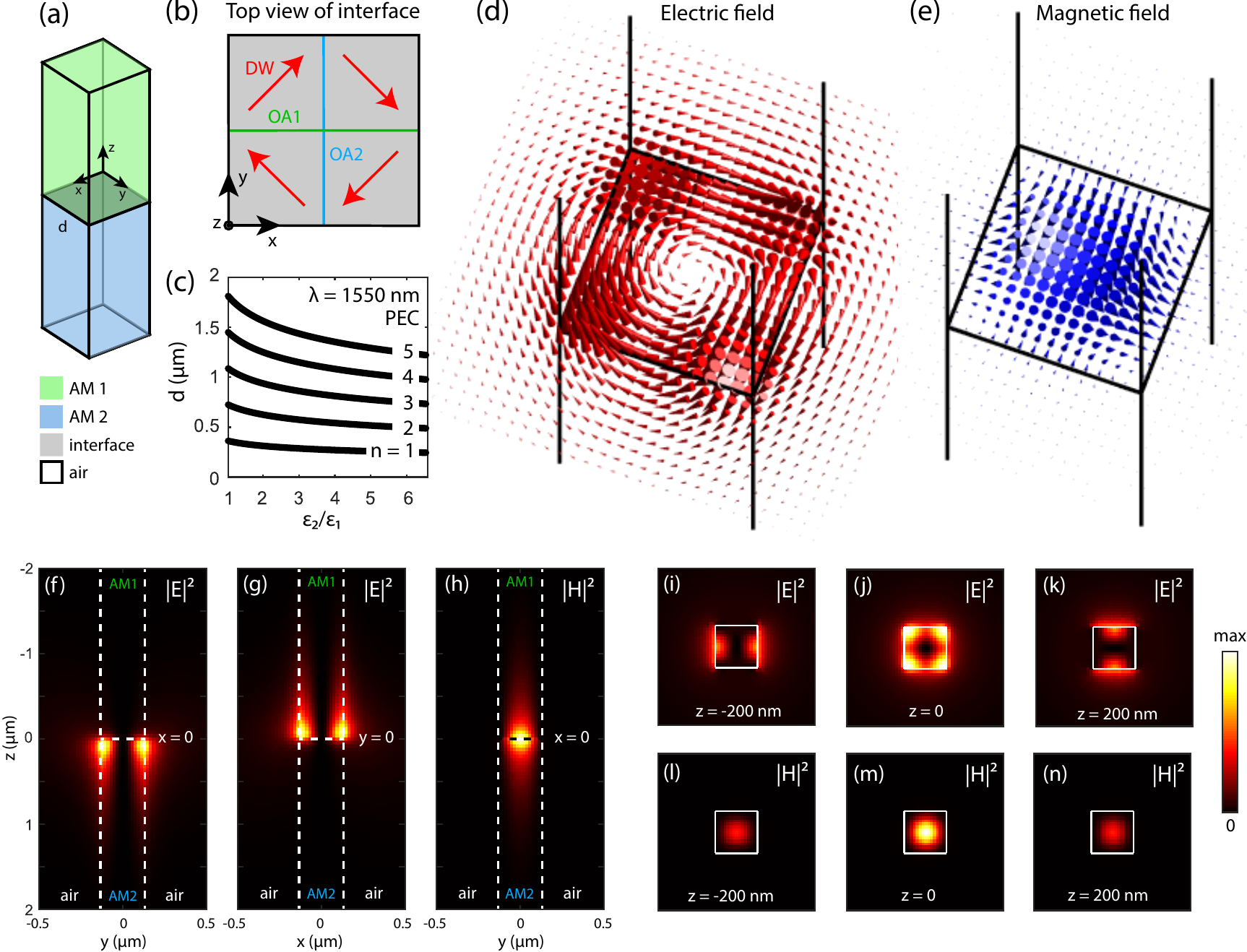}
    \caption{(a) The interface between two square cylinders made of anisotropic materials (AM). (b) The orientation of optical axes is shown by blue and green lines. Dyakonov waves reflect from the air boundaries at the angle of $\alpha = 45^\circ$ forming two-dimensionally confined mode. (c) The anisotropy factor dependence of square widths $d$ at which DSCM exist for different mode orders $n$. (d) Electric and (e) magnetic fields calculated in the horizontal plane $z=0$ in DSCM.  \textcolor{black}{Black lines denote the edges of rods.} Electric (f,g,i--k) and magnetic (h,l--n) field intensities in DSCM mode in horizontal (i--n) and vertical (f--h) cross-sections. $d=264.34$\,nm,  $\varepsilon_1=9$, $\varepsilon_2=36$, $\lambda=1550$\,nm.}
\label{fig:2d_confinement3}
\end{figure*}
\begin{figure*}[h]
    \centering
    \includegraphics[width = 0.9\textwidth]{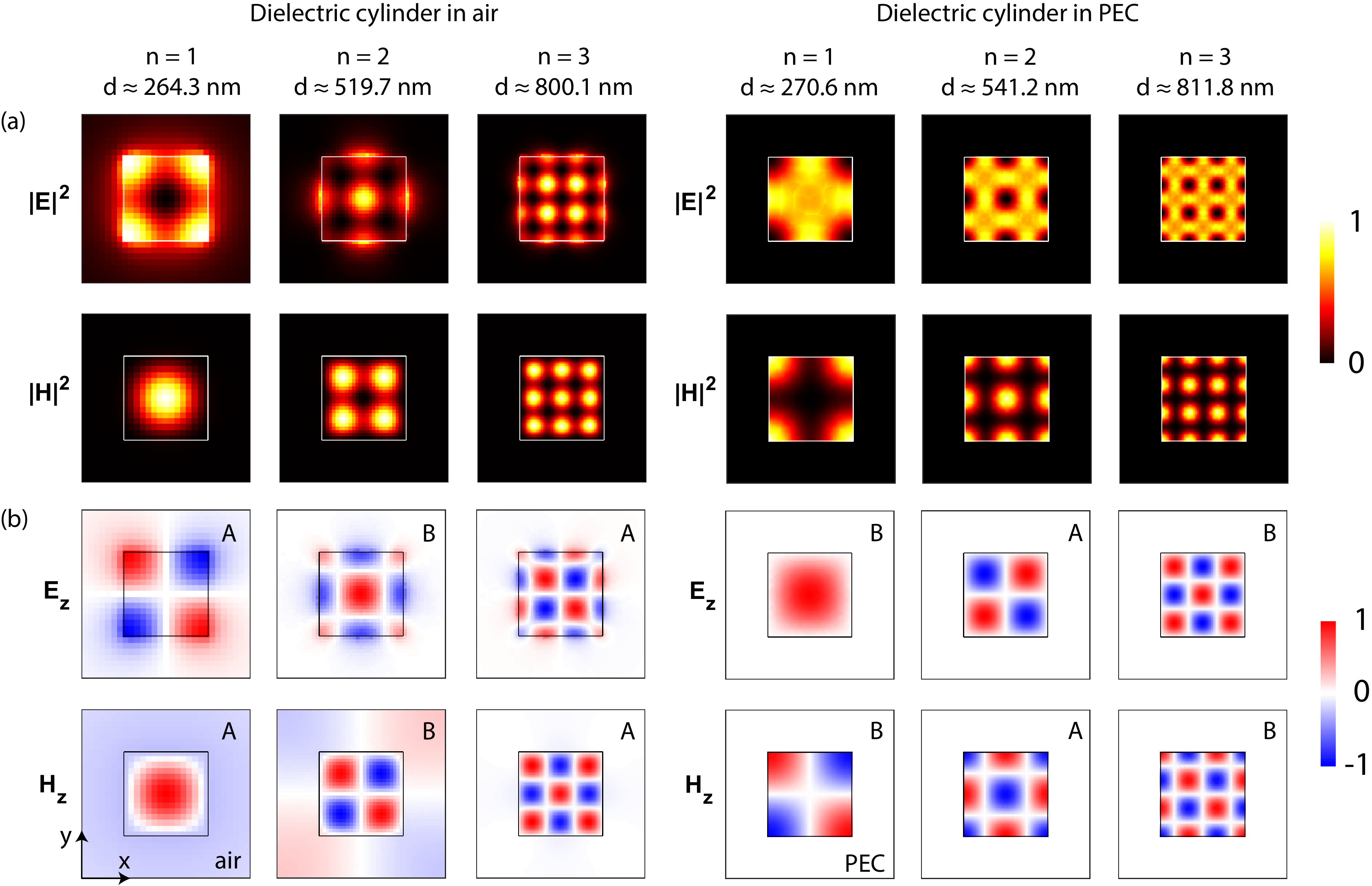}
    \caption{(Color online) Intensity (a) and $z$-projection (b) of electric and magnetic fields of DSCMs at $n=1-3$ calculated for the cases of air and PEC \textcolor{black}{boundaries}. Symbols A or B denote irreducible representations of the S$_4$ point group. $\varepsilon_1=9$, $\varepsilon_2=36$, $\lambda=1550$\,nm. Colorscales are shown on the right.}
\label{fig:2d_confinement2}
\end{figure*}
Due to symmetrical configuration of DSWs relative to the optical axes (Fig.\,\ref{fig:infinite_space}b), one can confine DSWs in two dimensions using two pairs of orthogonal boundaries, as shown in Figs.\,\ref{fig:2d_confinement3}a,b. In such a system, the DSW reflects at an angle of $\alpha=45^{\circ}$ from four boundaries forming a closed Dyakonov-like surface cavity mode (DSCM), which exists at the interface between two adjacent anisotropic rods of a square cross-section. In such configuration, optical axes of anisotropic materials have to be parallel to the square's sides. If the rods are surrounded by PEC then the DSCMs existence condition can be expressed by a simple formula:
\begin{align}
\begin{split}
    \beta(\pi/4) d/\sqrt{2}=\pi n,
    n\in\mathbb{N}
    \label{eq:2d_dispersion}
\end{split}
\end{align}
where $n$ is the mode order, $d$ is the side of the square, and $\beta(\pi/4)$ is the propagation constant of the DSW at the infinite interface determined by Eq. (\ref{beta}). Since this condition is only valid for $\alpha=\pi/4$, then, in contrast to one-dimensionally confined DSWMs, DSCMs exist only at a discrete set of the square side $d$  \textcolor{black}{(at a given frequency and dielectric permittivity)}. The calculated by Eq.\,\eqref{eq:2d_dispersion} values of the square side supporting DSCM are shown in Fig.\,\ref{fig:2d_confinement3}c as a function of the anisotropy factor for mode orders $n=$1--5 and PEC boundaries. These theoretical dependencies are confirmed by full-wave simulations in COMSOL Multiphysics. One can also obtain similar curves for the air boundaries.


We note that the structure with anisotropic rods shown in Figs.\,\ref{fig:2d_confinement3}a,b is S$_4$-symmetrical, i.e. it is invariant under 90$^{\circ}$ rotation about the $z$-axis and subsequent mirror reflection relative to the $xy$-plane. The group theory tells us that i) S$_4$-symmetrical structures should have two singlet eigenmodes and one doublet eigenmode; ii) symmetries of these eigenmodes are determined by irreducible representations of the S$_4$ point group. To explore this in application to our system consisting of two tangent rods surrounded by air or PEC, we simulate its eigenmodes in COMSOL. \textcolor{black}{Our simulations revel that, as expected, such} a system supports waveguide modes propagating along the rods, as well as DSCMs localized at the interface between the rods. We notice that the waveguide modes can be singlets or doublets while the DSCMs are always singlets. This is due to more strict selection rules for DSCM in comparison to waveguide modes. The calculated electric and magnetic field intensity profiles of the 1-st order DSCM in vertical and horizontal cross sections are shown in Figs.\,\ref{fig:2d_confinement3}f--n. One can see that in the upper (or lower) rod the electric field is mainly localized near the boundaries $y=\mathrm{const}$ (or $x=\mathrm{const}$), whereas at the interface between the rods ($z=0$) it is localized near the four square's corners, thereby, having a higher degree of symmetry. Also, at the interface the electric vector takes the vortex shape, while magnetic field is maximal in the center of the vortex (Fig.\,\ref{fig:2d_confinement3}d,e). \textcolor{black}{Such a structure of DSCM is similar to the Mie resonances presented in Ref.~\cite{Garcia-Etxarri2011}. Using these resonances, it is possible to obtain a strong magnetic response from a magnetic particle placed in the centre of the interface.}  Electric and magnetic field intensity profiles in the DSCMs of the orders $n=1$--3 are shown in Fig\,\ref{fig:2d_confinement2}a. We notice that due to different boundary conditions, the number of nodes (or antinodes) in electric (or magnetic) intensity field profile equals to $n\times n$ for the air \textcolor{black}{boundary}, and to $(n+1)\times(n+1)$ for the PEC \textcolor{black}{boundary}. To ascribe a specific irreducible representation to the obtained DSCMs, we also plot the $z$-projection of electric and magnetic vectors in Figs.\,\ref{fig:2d_confinement2}b. By inspecting Figs.\,\ref{fig:2d_confinement2}b one can see that the displayed DSCMs refer to the irreducible representations either A or B of the S$_4$ point group \footnote{The difference between irreducible representations A and B in S$_4$ point group is whether a field changes its sign under the symmetry operation S$_4$. See \,\citenum{symmetrylib} for details.}.

Due to radiation losses caused by scattering of DSWs at the air boundaries, DSCMs should have a finite Q-factor when rods are surrounded by air. Generally, the Q-factor depends on the dielectric permittivities of rods and environment, as well as on the mode order. The COMSOL simulation reveals that for the case of air boundaries and components of dielectric tensors of rods $\varepsilon_1=9$ and \textcolor{black}{$\varepsilon_{2}=36$}, the Q-factor of DSCMs with $n=$1, 2 and 3 equal to 8.83, 52.98 and 197.06, respectively. \textcolor{black}{Apparently, the Q-factor decreases with the mode order due to increasing diffraction losses caused by the violation of the total internal reflection in smaller squares.} Due to lack of scattering and absorption losses in the rods surrounded by PEC, the corresponding Q-factor is infinite. 

\textcolor{black}{Finally, DSCMs demonstrated in this section, can be generalized to the case of cylinders of an arbitrary rectangular shape. A DSCM in such a structure is a superposition of DSWs reflecting from boundaries at angles $\alpha$ not equal to 45 degrees and propagating at azimuthal angles $\varphi$ also not equal to 45 degrees. Such a structure is no longer S$_4$-symmetrical and the corresponding field distributions are less symmetrical in comparison to the square cylinders case (See Supplemental Materials for details). Due to the perfect reflection from the PEC boundary (Fig.\,\ref{fig:reflection_case}d), the Q-factors of DSCMs in rectangular cylinders with PEC boundaries remain infinite.}

\section{Conclusion}
In conclusion, we have studied Dyakonov-like surface states which appear at the interface between two identical anisotropic dielectrics \textcolor{black}{twisted in such a way that their}  optical axes form an angle of 90$^\circ$ to each other. First, we have studied the case of the infinite horizontal interface where Dyakonov surface waves exist in a small range of azimuthal angles. In the presence of vertical boundaries that constrain the system from two sides, electromagnetic confinement comes into play. We have demonstrated that such a one-dimensionally \textcolor{black}{confined} system supports Dyakonov surface waveguide modes propagating along the direction where conventional Dyakonov surface waves do not exist. We have shown that the 1-st order Dyakonov surface waveguide mode can propagate without losses. \textcolor{black}{This fact opens ample opportunities of using these modes in signal transmission lines and information processing.} The existence of Dyakonov surface waveguide modes can be explained in terms of the multireflection of Dyakonov surface waves at \textcolor{black}{angles close to 45$^\circ$} to the interfacial strip waveguide boundaries. We have further improved this idea and considered the interface between two square cylinders made of anisotropic materials. Due to the two-dimensional electromagnetic confinement, such a system supports Dyakonov surface cavity modes. We believe that our work can open new insights in the field of surface waves in anisotropic media, which can lead to the practical application of Dyakonov surface waves in optoelectronic devices.

\section{Theoretical methods}
To simulate the DSW reflection from a single boundary, we developed a model in COMSOL Multiphysics where the DSW is excited by a port plane. The field and the wavevector of the mode which are excited by the port are taken as a DSW solution at the infinite interface described in Sec.\,\textbf{Interface of two uniaxial crystals}. Then, we find the S-parameters of such a system by calculating the fields at the reflection and transmission sides. As a result, we obtain the total reflectance and transmittance of DSW at the boundary. \textcolor{black}{We verified our numerical results obtained in COMSOL Multiphysics using the analytical solutions. When calculating models that do not have an analytical solution, we checked that the final results do not depend on the grid size and the position of the PML layers.}

\begin{acknowledgement}
 Authors acknowledge Ilia M. Fradkin for fruitful discussions.
\end{acknowledgement}

\begin{funding}
  This work was supported by the Russian Foundation for Basic Research (Grant \textnumero 18-29-20032).
\end{funding}


\begin{thebibliography}{43}

\bibitem{Raether1988}
H.~Raether, ``{Intro{\_}Contents},'' {\em Intro{\_}Surface Plasmons on Smooth
  and Rough Surfaces and on Gratings}, p.~78, 1988.

\bibitem{Malkova2006a}
N.~Malkova and C.~Z. Ning, ``{Shockley and {T}amm surface states in photonic
  crystals},'' {\em Phys. Rev. B}, vol.~73, no.~11, p.~113113, 2006.

\bibitem{vinogradov10}
A.~P. Vinogradov, A.~V. Dorofeenko, A.~M. Merzlikin, and A.~A. Lisyansky,
  ``{Surface states in photonic crystals},'' {\em Physics-Uspekhi}, vol.~53,
  p.~243, 2010.

\bibitem{dyakov2012surface}
S.~A. Dyakov, A.~Baldycheva, T.~S. Perova, G.~V. Li, E.~V. Astrova, N.~A.
  Gippius, and S.~G. Tikhodeev, ``Surface states in the optical spectra of
  two-dimensional photonic crystals with various surface terminations,'' {\em
  Phys. Rev. B}, vol.~86, p.~115126, Sep 2012.

\bibitem{kartashov2006surface}
Y.~V. Kartashov, V.~A. Vysloukh, and L.~Torner, ``Surface gap solitons,'' {\em
  Physical review letters}, vol.~96, no.~7, p.~073901, 2006.

\bibitem{Dyakonov1988}
M.~I. D'yakonov, ``{New type of electromagnetic wave propagating at an
  interface},'' {\em Sov Phys JETP}, vol.~67, no.~April, pp.~714--716, 1988.

\bibitem{Walker1998}
D.~B. Walker, E.~N. Glytsis, and T.~K. Gaylord, ``{Surface mode at
  isotropic–uniaxial and isotropic–biaxial interfaces},'' {\em Journal of
  the Optical Society of America A}, vol.~15, no.~1, p.~248, 1998.

\bibitem{Narimanov2018}
E.~E. Narimanov, ``{Dyakonov waves in biaxial anisotropic crystals},'' {\em
  Physical Review A}, vol.~98, no.~1, pp.~1--13, 2018.

\bibitem{Lakhtakia2007}
A.~Lakhtakia and J.~A. Polo, ``{Dyakonov-Tamm wave at the planar interface of a
  chiral sculptured thin film and an isotropic dielectric material},'' {\em
  Journal of the European Optical Society}, vol.~2, pp.~1--12, 2007.

\bibitem{Gao:09}
J.~Gao, A.~Lakhtakia, and M.~Lei, ``On dyakonov-tamm waves localized to a
  central twist defect in a structurally chiral material,'' {\em J. Opt. Soc.
  Am. B}, vol.~26, pp.~B74--B82, Dec 2009.

\bibitem{repan2020wave}
T.~Rep{\"a}n, O.~Takayama, and A.~V. Lavrinenko, ``Wave front tuning of coupled
  hyperbolic surface waves on anisotropic interfaces,'' in {\em Photonics},
  vol.~7, p.~34, Multidisciplinary Digital Publishing Institute, 2020.

\bibitem{zhang2020unusual}
Y.~Zhang, X.~Wang, D.~Zhang, S.~Fu, S.~Zhou, and X.-Z. Wang, ``Unusual spin and
  angular momentum of dyakonov waves at the hyperbolic-material surface,'' {\em
  Optics Express}, vol.~28, no.~13, pp.~19205--19217, 2020.

\bibitem{karpov2019dyakonov}
S.~Y. Karpov, ``Dyakonov surface electromagnetic waves in iii-nitride
  heterostructures,'' {\em physica status solidi (b)}, vol.~256, no.~3,
  p.~1800609, 2019.

\bibitem{fedorin2019dyakonov}
I.~Fedorin, ``Dyakonov surface waves at the interface of nanocomposites with
  spherical and ellipsoidal inclusions,'' {\em Optical and Quantum
  Electronics}, vol.~51, no.~6, p.~201, 2019.

\bibitem{Takayama2008}
O.~Takayama, L.~C. Crasovan, S.~K. Johansen, D.~Mihalache, D.~Artigas, and
  L.~Torner, ``{Dyakonov surface waves: A review},'' {\em Electromagnetics},
  vol.~28, no.~3, pp.~126--145, 2008.

\bibitem{Takayama2009}
O.~Takayama, L.~Crasovan, D.~Artigas, and L.~Torner, ``{Observation of dyakonov
  surface waves},'' {\em Physical Review Letters}, vol.~102, no.~4, pp.~2--5,
  2009.

\bibitem{Takayama2014}
O.~Takayama, D.~Artigas, and L.~Torner, ``{Lossless directional guiding of
  light in dielectric nanosheets using Dyakonov surface waves},'' {\em Nature
  Nanotechnology}, vol.~9, no.~6, pp.~419--424, 2014.

\bibitem{Chiadini2016}
F.~Chiadini, V.~Fiumara, A.~Scaglione, and A.~Lakhtakia, ``{Compound guided
  waves that mix characteristics of surface-plasmon-polariton, Tamm,
  Dyakonov–Tamm, and Uller–Zenneck waves},'' {\em Journal of the Optical
  Society of America B}, vol.~33, no.~6, p.~1197, 2016.

\bibitem{Lakhtakia2014}
A.~Lakhtakia and M.~Faryad, ``{Theory of optical sensing with Dyakonov–Tamm
  waves},'' {\em Journal of Nanophotonics}, vol.~8, p.~083072, nov 2014.

\bibitem{Pulsifer2013}
D.~P. Pulsifer, M.~Faryad, and A.~Lakhtakia, ``{Observation of the
  Dyakonov-Tamm wave},'' {\em Physical Review Letters}, vol.~111, no.~24,
  pp.~1--5, 2013.

\bibitem{artigas2005dyakonov}
D.~Artigas and L.~Torner, ``Dyakonov surface waves in photonic metamaterials,''
  {\em Physical review letters}, vol.~94, no.~1, p.~013901, 2005.

\bibitem{jacob2008optical}
Z.~Jacob and E.~E. Narimanov, ``Optical hyperspace for plasmons: Dyakonov
  states in metamaterials,'' {\em Applied Physics Letters}, vol.~93, no.~22,
  p.~221109, 2008.

\bibitem{Takayama2012}
O.~Takayama, D.~Artigas, and L.~Torner, ``{Practical dyakonons},'' {\em Optics
  Letters}, vol.~37, no.~20, p.~4311, 2012.

\bibitem{Takayama2012coupling}
O.~Takayama, D.~Artigas, and L.~Torner, ``{Coupling plasmons and dyakonons},''
  {\em Optics Letters}, vol.~37, p.~1983, jun 2012.

\bibitem{Mackay2019}
T.~G. Mackay, C.~Zhou, and A.~Lakhtakia, ``{Dyakonov–Voigt surface waves},''
  {\em Proceedings of the Royal Society A: Mathematical, Physical and
  Engineering Sciences}, vol.~475, p.~20190317, aug 2019.

\bibitem{Lakhtakia2020}
A.~Lakhtakia and T.~Mackay, ``{From unexceptional to doubly exceptional surface
  waves},'' {\em Journal of the Optical Society of America B}, no.~2,
  pp.~1--13, 2020.

\bibitem{Lakhtakia2020a}
A.~Lakhtakia, T.~G. Mackay, and C.~Zhou, ``{Electromagnetic surface waves at
  exceptional points},'' pp.~1--9, 2020.

\bibitem{zhou2020theory}
C.~Zhou, T.~G. Mackay, and A.~Lakhtakia, ``Theory of dyakonov--tamm surface
  waves featuring dyakonov--tamm--voigt surface waves,'' {\em Optik},
  p.~164575, 2020.

\bibitem{Kajorndejnukul2019}
V.~Kajorndejnukul, D.~Artigas, and L.~Torner, ``{Conformal transformation of
  Dyakonov surface waves into bound states of cylindrical metamaterials},''
  {\em Physical Review B}, vol.~100, no.~19, pp.~1--6, 2019.

\bibitem{Golenitskii2019}
K.~Y. Golenitskii and A.~A. Bogdanov, ``{Dyakonov-like surface waves in
  anisotropic cylindrical waveguides},'' 2019.

\bibitem{averkiev1990electromagnetic}
N.~Averkiev and M.~Dyakonov, ``Electromagnetic waves localized at the boundary
  of transparent anisotropic media,'' {\em Optics and Spectroscopy}, vol.~68,
  no.~5, pp.~1118--1121, 1990.

\bibitem{Darinskii2001}
A.~N. Darinskii, ``{Dispersionless polaritons on a twist boundary in optically
  uniaxial crystals},'' {\em Crystallography Reports}, vol.~46, no.~5,
  pp.~842--844, 2001.

\bibitem{Furs2005}
A.~N. Furs, V.~M. Galynsky, and L.~M. Barkovsky, ``{Dispersionless surface
  polaritons at twist boundaries of crystals and in a transition layer between
  the crystals},'' {\em Optics and Spectroscopy}, vol.~98, no.~3, pp.~454--461,
  2005.

\bibitem{Furs2005a}
A.~N. Furs, V.~M. Galynsky, and L.~M. Barkovsky, ``{Surface polaritons in
  symmetry planes of biaxial crystals},'' {\em Journal of Physics A:
  Mathematical and General}, vol.~38, pp.~8083--8101, sep 2005.

\bibitem{Nelatury2007}
S.~R. Nelatury, J.~A. {Polo, Jr.}, and A.~Lakhtakia, ``{Surface waves with
  simple exponential transverse decay at a biaxial bicrystalline interface},''
  {\em Journal of the Optical Society of America A}, vol.~24, p.~856, mar 2007.

\bibitem{PoloJr.2007}
J.~A. {Polo, Jr.}, S.~R. Nelatury, and A.~Lakhtakia, ``{Surface waves at a
  biaxial bicrystalline interface},'' {\em Journal of the Optical Society of
  America A}, vol.~24, no.~9, p.~2974, 2007.

\bibitem{Alshits2002}
V.~I. Alshits and V.~N. Lyubimov, ``{Dispersionless surface polaritons in the
  vicinity of different sections of optically uniaxial crystals},'' {\em
  Physics of the Solid State}, vol.~44, no.~2, pp.~386--390, 2002.

\bibitem{Takayama2011}
O.~Takayama, A.~Y. Nikitin, L.~Martin-Moreno, L.~Torner, and D.~Artigas,
  ``{Dyakonov surface wave resonant transmission},'' {\em Optics Express},
  vol.~19, no.~7, p.~6339, 2011.

\bibitem{li2003fourier}
L.~Li, ``Fourier modal method for crossed anisotropic gratings with arbitrary
  permittivity and permeability tensors,'' {\em Journal of Optics A: Pure and
  Applied Optics}, vol.~5, no.~4, p.~345, 2003.

\bibitem{weiss2009matched}
T.~Weiss, G.~Granet, N.~A. Gippius, S.~G. Tikhodeev, and H.~Giessen, ``Matched
  coordinates and adaptive spatial resolution in the fourier modal method,''
  {\em Optics express}, vol.~17, no.~10, pp.~8051--8061, 2009.

\bibitem{Tikhodeev2002b}
S.~G. Tikhodeev, A.~L. Yablonskii, E.~A. Muljarov, N.~A. Gippius, and
  T.~Ishihara, ``Quasiguided modes and optical properties of photonic crystal
  slabs,'' {\em Phys. Rev. B}, vol.~66, p.~045102, 2002.

\bibitem{Garcia-Etxarri2011}
A.~Garc{\'{i}}a-Etxarri, R.~G{\'{o}}mez-Medina, L.~S. Froufe-P{\'{e}}rez,
  C.~L{\'{o}}pez, L.~Chantada, F.~Scheffold, J.~Aizpurua, M.~Nieto-Vesperinas,
  and J.~J. S{\'{a}}enz, ``{Strong magnetic response of submicron Silicon
  particles in the infrared},'' {\em Optics Express}, vol.~19, p.~4815, mar
  2011.

\bibitem{symmetrylib}
\url{http://symmetry.jacobs-university.de}.

\end{thebibliography}

\end{document}